\documentclass[aps,prb,preprint,groupedaddress,floatfix,onecolumn]{revtex4-2}

\usepackage{graphicx}
\usepackage{xr}
\usepackage{amsmath}
 \usepackage{amssymb}
 \usepackage{amsfonts}
 \usepackage{color}

\begin{document}


\title{Neural-network-based reconstruction of spin and orbital angular momentum from X-ray magnetic circular dichroism spectra}


\author{Tetsuro Ueno}
\email[Contact author: ]{ueno.tetsuro@qst.go.jp}
\affiliation{Synchrotron Radiation Research Center, Kansai Institute for Photon Science, National Institutes for Quantum Science and Technology, Sayo 679-5148, Japan}
\affiliation{Quantum Materials and Applications Research Center, Takasaki Institute for Advanced Quantum Science, National Institutes for Quantum Science and Technology, Takasaki 370-1292, Japan}


\date{\today}

\begin{abstract}
X-ray magnetic circular dichroism (XMCD) is a powerful probe of element-specific spin and orbital angular momentum. 
Conventional analyses based on sum rules, however, rely on integrated spectral intensities and can become insufficient when multiple parameters influence the spectral line shape. 
Here, we formulate XMCD analysis as an inverse problem and develop a neural-network (NN) based approach to reconstruct spin and orbital angular momentum directly from full spectral line shapes. 
Using many-body multiplet calculations of Fe, Co, and Ni $L_{2,3}$-edge X-ray absorption spectra (XAS) and XMCD spectra as a physically well-defined training dataset, we systematically vary key parameters including crystal-field splitting, spin--orbit coupling, and exchange field. 
The NN is trained to map spectral line shapes onto the expectation values of spin and orbital angular momentm $\langle S_z \rangle$ and $\langle L_z \rangle$, and validated using strictly test-only data. 
The results demonstrate accurate and unbiased reconstruction, establishing a proof of concept for data-driven inverse reconstruction from XAS and XMCD spectra. 
These findings show that exploiting the full XAS and XMCD line shapes provide access to information beyond conventional sum-rule analyses while remaining consistent with established theoretical frameworks.
\end{abstract}


\maketitle

\section{Introduction}
\label{lntroduction}

X-ray magnetic circular dichroism (XMCD) spectroscopy is a powerful technique for probing element-specific magnetic properties of materials, providing direct sensitivity to spin and orbital angular momentum through polarization-dependent X-ray absorption processes~\cite{vanderLaan_2014, Vaz_2025}. 
As a magnetic extension of X-ray absorption spectroscopy (XAS), XMCD has been widely applied to magnetic materials and extended to spatially resolved, time-resolved, and operando measurements~\cite{Fischer_2017, Wadati_2026, Miyakaze_2020}. 
Quantitative analysis is commonly based on magneto-optical sum rules, which relate energy-integrated signals to spin and orbital magnetic moments~\cite{Thole_1992, Carra_1993, Chen_1995, vanderLaan_1999}.

Despite its broad applicability, quantitative XMCD analysis remains challenging. 
Sum-rule-based approaches rely on accurate background subtraction and normalization, which are often empirical and operator dependent, limiting reproducibility~\cite{Yokoyama_2025}. 
These challenges are particularly severe for weak signals, such as those obtained in total electron yield measurements or dilute systems.

Recent advances in autonomous experimentation and high-throughput XMCD measurements have further increased the demand for reliable and rapid data analysis~\cite{Ishizuki_2023, Yoshikawa_2025, Yamazaki_2025, Yamasaki_2025}. 
In such settings, manual analysis becomes a bottleneck and the extraction of angular-momentum expectation values often requires substantial expertise.

In this context, machine-learning (ML) techniques have emerged as powerful tools for spectral analysis~\cite{Ueno_2022, Aoyagi_2023, Penfold_2024, Westermayr_2025}. 
Data-driven approaches have been successfully applied to XAS and related spectroscopies to extract structural and electronic information~\cite{Timoshenko_2018, Han_2025, Guda_2021}. 
In particular, neural networks (NNs) can capture complex nonlinear relationships between spectral features and physical quantities, enabling efficient analysis beyond conventional fitting-based methods~\cite{Haykin_NNbook}.

From this perspective, XMCD analysis can be viewed as an inverse problem. 
While the forward calculation of XMCD spectra is well established, extracting spin and orbital angular momentum from full spectral line shapes remains highly non-trivial. 
Conventional sum rules exploit only integrated intensities, whereas the full spectra encode rich information on crystal-field splitting, spin--orbit coupling, and exchange interactions. 
However, a systematic framework to utilize this information has been lacking.

In this work, we formulate XMCD analysis as an inverse problem and develop a NN-based approach to reconstruct spin and orbital angular momentum directly from full XAS and XMCD spectral line shapes. 
A physically well-defined dataset is constructed using many-body calculations of typical ferromagnetic $3d$ elements, Fe, Co, and Ni $L_{2,3}$ XAS and XMCD spectra with systematically varied parameters, and the NN is trained to map spectral features onto the expectation values of spin and orbital angular momentm $\langle S_z \rangle$ and $\langle L_z \rangle$. 
The results demonstrate accurate and unbiased reconstruction, establishing a proof of concept for data-driven inverse analysis of XMCD spectra.

This paper is organized as follows. 
In Sec.~\ref{Methods}, we describe the workflow, dataset construction, and NN model. 
Section~\ref{Results} presents the results, followed by discussion in Sec.~\ref{Discussion}. 
Conclusions are given in Sec.~\ref{Conclusion}.

\section{Methods}
\label{Methods}

\subsection{Workflow}

\begin{figure*}[tb]
\includegraphics[width=\linewidth]{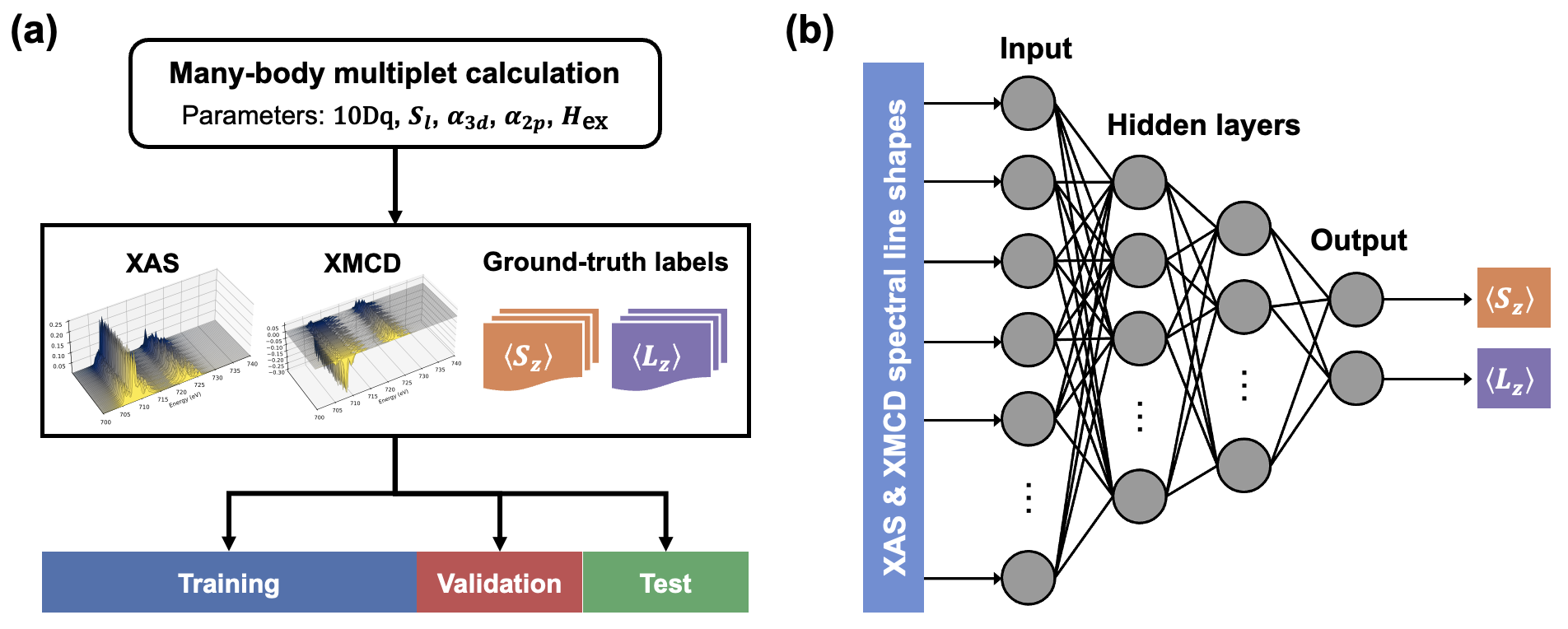}%
\caption{\label{Fig:concept}
Workflow of the NN-based reconstruction of spin and orbital angular-momentum expectation values.
(a) Dataset generated from many-body multiplet calculations, including XAS and XMCD spectra together with the corresponding ground-truth labels $\langle S_z \rangle$ and $\langle L_z \rangle$, obtained by systematically varying physical parameters. 
The dataset, excluding data points with $\langle S_z \rangle = 0$ and $\langle L_z \rangle = 0$, is divided into training, validation, and test subsets.
(b) NN architecture used in this study.
Filled circles and solid lines represent nodes and edges of NN, respectively.
The XAS and XMCD spectral line shapes are used as inputs to fully connected hidden layers with a shared representation, followed by two output nodes corresponding to $\langle S_z \rangle$ and $\langle L_z \rangle$.
}
\end{figure*}

Figure~\ref{Fig:concept} schematically illustrates the overall workflow of the present approach, which combines many-body calculations of XAS and XMCD spectra with an NN to formulate XMCD analysis as an inverse problem.
First, many-body multiplet calculations are performed to generate a dataset consisting of XAS and XMCD spectra together with the corresponding expectation values of $\langle S_z \rangle$ and $\langle L_z \rangle$, obtained by systematically varying physical parameters. 
Data points with $\langle S_z \rangle = 0$ and $\langle L_z \rangle = 0$ are excluded, and the remaining dataset is randomly divided into training, validation, and test subsets, as shown in Fig.~\ref{Fig:concept}(a).
Next, the NN is trained to learn the inverse mapping from the spectral data to the angular-momentum expectation values. 
The trained model is then used to reconstruct spin and orbital angular momentum directly from XAS and XMCD spectra, as illustrated in Fig.~\ref{Fig:concept}(b). 
Through this workflow, XMCD analysis is cast as a well-defined inverse problem that maps spectral features onto microscopic magnetic observables.
To explicitly formulate this inverse problem, we describe the mapping from spectral line shapes to the angular-momentum expectation values as a parametric function learned by a neural network.
The input consists of concatenated XAS and XMCD spectra, while the output corresponds to the spin and orbital angular momentum expectation values.
This relationship is expressed as
\begin{equation}
\mathbf{y} = f_\theta(\mathbf{x}),
\end{equation}
where $\mathbf{x} \in \mathbb{R}^{2N_E}$ denotes the discretized XAS and XMCD spectra, and $\mathbf{y} = \left( \langle S_z \rangle, \langle L_z \rangle \right) \in \mathbb{R}^2$ represents the target quantities.
Here, $f_\theta$ denotes the neural network with trainable parameters $\theta$.
Details of each step are provided in the following sections.

\subsection{Many-body multiplet calculations of XAS and XMCD spectra}

To generate a physically well-defined dataset for training and validation of the NN, XAS and XMCD spectra were calculated using the Quanty code, which is based on full-multiplet many-body theory~\cite{Haverkort_2016}.
We calculated XAS and XMCD spectra of Fe, Co, and Ni $L_{2,3}$ absorption edges, corresponding to dipole-allowed $2p$--$3d$ transitions.
The $3d$ electron configuration is assumed to be $3d^6$, $3d^7$ and $3d^8$ for Fe, Co, and Ni, respectively, and the many-body Hamiltonian includes the essential interactions required to describe the $2p$--$3d$ excitation process.
Specifically, we take into account the crystal-field splitting parameterized by 10Dq, the $2p$--$3d$ and $3d$--$3d$ Coulomb interactions expressed in terms of
Slater integrals scaled by a factor $S_l$, the spin--orbit coupling for both the $2p$ core level and the $3d$ valence shell, and an effective exchange field
$H_{\mathrm{ex}}$ acting on the $3d$ electrons.
The spin--orbit coupling strengths are expressed as dimensionless scaling factors relative to the corresponding atomic Hartree--Fock values, $\zeta_{nl} = \alpha_{nl}\,\zeta_{nl}^{\mathrm{HF}}$ (with $nl = 2p, 3d$), following the standard implementation in Quanty~\cite{Ackermann_2024}.
These terms jointly determine the detailed XMCD spectral line shapes through multiplet effects and spin-dependent selection rules.

The calculation parameters were systematically varied within physically relevant ranges to sample a broad range of spectral configurations.
The multiplet calculation parameters are summarized in Table~\ref{Table:multiplet_parameters}.
The temperature was fixed at 10~K, and an external magnetic field of 2~T was applied along the incident X-ray propagation direction in all calculations.
For each parameter set, the ground-state expectation values of the spin and orbital angular momentum, $\langle S_z \rangle$ and $\langle L_z \rangle$, were evaluated directly from the many-body wave functions and used as the corresponding target labels.
By constructing the dataset in this manner, a one-to-one correspondence between the XAS and XMCD spectra and physically meaningful angular-momentum expectation values is established, enabling a rigorous formulation of XMCD analysis as an inverse problem suitable for supervised machine learning.

\begin{table*}[tb]
\caption{\label{Table:multiplet_parameters}
Parameter ranges used in the many-body multiplet calculations for generating the dataset.
The spin--orbit coupling strengths $\alpha_{3d}$ and $\alpha_{2p}$ are expressed as dimensionless scaling factors relative to the corresponding atomic Hartree--Fock values $\zeta_{3d}^{\mathrm{HF}}$ and $\zeta_{2p}^{\mathrm{HF}}$.}
\centering
\begin{ruledtabular}
\begin{tabular}{llll}
\textbf{Parameter} & \textbf{Symbol} & \textbf{Range} & \textbf{Description} \\
\hline
Crystal-field splitting
& $10$Dq
& $0.8$--$1.6~\mathrm{eV}$
& Octahedral crystal-field strength \\

Slater-integral scaling
& $S_l$
& $0.7$--$0.9$
& Reduction factor for $3d$--$3d$ and $2p$--$3d$ Coulomb integrals \\

Spin-orbit coupling (3$d$)
& $\alpha_{3d}$
& $0.7$--$1.3$
& Valence-shell spin-orbit coupling \\

Spin-orbit coupling (2$p$)
& $\alpha_{2p}$
& $0.9$--$1.1$
& Core-level spin-orbit coupling \\

Exchange field
& $H_{\mathrm{ex}}$
& $0$--$0.01~\mathrm{eV}$
& Effective exchange field acting on $3d$ electrons \\
\end{tabular}
\end{ruledtabular}
\end{table*}

\subsection{Dataset preprocessing}

The input to the NN consists of XAS and XMCD spectra calculated for each parameter set, while the output targets are $\langle S_z \rangle$ and $\langle L_z \rangle$. 
Each spectrum is represented as a one-dimensional vector sampled on a fixed energy grid, ensuring a consistent input dimensionality.
Prior to training, the input spectra are normalized to remove trivial scale factors and emphasize the spectral shape. 
To improve numerical stability and facilitate efficient training, the input spectra and target values are normalized using standard score normalization,
\begin{equation}
\tilde{\mathbf{x}} = \frac{\mathbf{x} - \boldsymbol{\mu}_x}{\boldsymbol{\sigma}_x},
\end{equation}
\begin{equation}
\tilde{\mathbf{y}} = \frac{\mathbf{y} - \boldsymbol{\mu}_y}{\boldsymbol{\sigma}_y},
\end{equation}
where $\boldsymbol{\mu}$ and $\boldsymbol{\sigma}$ denote the mean and standard deviation of the dataset, respectively.
The normalized variables are used during training, and the predictions are transformed back to physical units for evaluation.
The target quantities are also normalized during training to facilitate numerical stability, and are converted back to physical units after evaluation.
Data points with $\langle S_z \rangle = 0$ and $\langle L_z \rangle = 0$ are excluded, as they correspond to vanishing XMCD signals and do not contribute to learning the inverse mapping. 
The dataset is randomly divided into mutually exclusive training, validation, and test subsets. 
The dataset consists of 3,240 samples, split into 2,268 training, 486 validation, and 486 test samples for each element.

\subsection{Neural-network architecture}

The NN is designed to perform supervised regression from XAS and XMCD spectra to physically meaningful angular-momentum expectation values. 
The network is implemented using the Python library PyTorch~\cite{PyTorch}. 
The input layer consists of neurons corresponding to the discretized XAS and XMCD spectra, with the total number of input neurons equal to twice the number of energy points used to represent each spectrum.
This representation allows the network to directly access the full spectral line shapes without relying on any hand-crafted feature extraction.

The input layer is followed by a shared feature extractor composed of fully connected hidden layers with 256 and 128 nodes.
The forward propagation of the network can be expressed as
\begin{align}
\mathbf{h}^{(1)} &= \sigma(\mathbf{W}_1 \mathbf{x} + \mathbf{b}_1), \\
\mathbf{h}^{(2)} &= \sigma(\mathbf{W}_2 \mathbf{h}^{(1)} + \mathbf{b}_2), \\
\mathbf{y} &= \mathbf{W}_3 \mathbf{h}^{(2)} + \mathbf{b}_3,
\end{align}
where $\mathbf{h}^{(1)}$ and $\mathbf{h}^{(2)}$ are hidden representations, and $\sigma(x) = \max(0, x)$ denotes the rectified linear unit (ReLU) activation function.
The output of these shared hidden layers forms a common latent representation of the XAS and XMCD spectral line shapes, capturing nonlinear relationships between spectral features and the underlying physical quantities. 
All hidden layers employ ReLU functions to introduce nonlinearity, together with batch normalization and dropout to stabilize training and mitigate overfitting~\cite{ReLU, BatchNorm, Dropout}. 
The depth and width of the network are chosen to balance model capacity and numerical stability while retaining sufficient flexibility to represent the inverse mapping.
The shared representation is mapped to two independent output heads, each consisting of a single linear neuron corresponding to $\langle S_z \rangle$ and $\langle L_z \rangle$, respectively. 
A linear activation is used in the output layers to allow unrestricted regression of these continuous physical quantities. 
This multi-output architecture enables the network to learn a direct and unified mapping from XAS and XMCD spectra to angular-momentum expectation values.
The detailed network architecture is summarized in Table~\ref{Table:nn_architecture}.

\begin{table*}[tb]
\caption{
Detailed architecture of the neural network.
Batch normalization (BatchNorm) is applied to the hidden layers to improve numerical stability and training convergence.
Rectified linear unit (ReLU) functions are used to introduce nonlinearity.
Dropout ($p=0.1$) is employed during training to mitigate overfitting.
}
\label{Table:nn_architecture}
\centering
\begin{ruledtabular}
\begin{tabular}{lll}
\textbf{Layer} & \textbf{Type / Operation} & \textbf{Dimension} \\
\hline
Input
& Discretized XAS and XMCD spectra
& $2N_E$ \\

Hidden layer 1
& Fully connected + BatchNorm + ReLU + Dropout
& 256 \\

Hidden layer 2
& Fully connected + BatchNorm + ReLU
& 128 \\

Output heads ($\langle S_z \rangle$, $\langle L_z \rangle$)
& Fully connected + linear
& 2 \\

\end{tabular}
\end{ruledtabular}
\end{table*}

\subsection{Training procedure}

The NN is trained in a supervised manner by minimizing the mean squared error (MSE) between the predicted and target values of $\langle S_z \rangle$ and $\langle L_z \rangle$.
The loss function is defined as
\begin{equation}
\mathcal{L}(\theta) =
\frac{1}{N} \sum_{i=1}^{N}
\left\| f_\theta(\mathbf{x}_i) - \mathbf{y}_i \right\|^2,
\end{equation}
where $N$ is the number of training samples.
In terms of the individual components, the loss can be written as
\begin{equation}
\mathcal{L}(\theta) =
\frac{1}{N} \sum_{i=1}^{N}
\left[
\left( \hat{S}_{z,i} - S_{z,i} \right)^2
+
\left( \hat{L}_{z,i} - L_{z,i} \right)^2
\right],
\end{equation}
where $\hat{S}_z$ and $\hat{L}_z$ denote the predicted values.
This loss function provides a natural measure of regression accuracy for continuous physical quantities and enables stable optimization of the network parameters.

Model optimization is performed using a gradient-based optimizer, which efficiently updates the network weights based on the back propagation of the loss function~\cite{Rumelhart_1986}.
The learning rate and related optimizer parameters are chosen to ensure numerical stability and reliable convergence during training.
To prevent overfitting and to improve generalization performance, early stopping is employed based on the validation loss, and the training process is terminated when no further improvement is observed over a predefined number of epochs.
Key hyperparameters, including the number of hidden layers, the number of neurons per layer, the learning rate, and the early-stopping criteria, are determined through validation-based tuning.

\begin{figure*}[h]
\includegraphics[width=\linewidth]{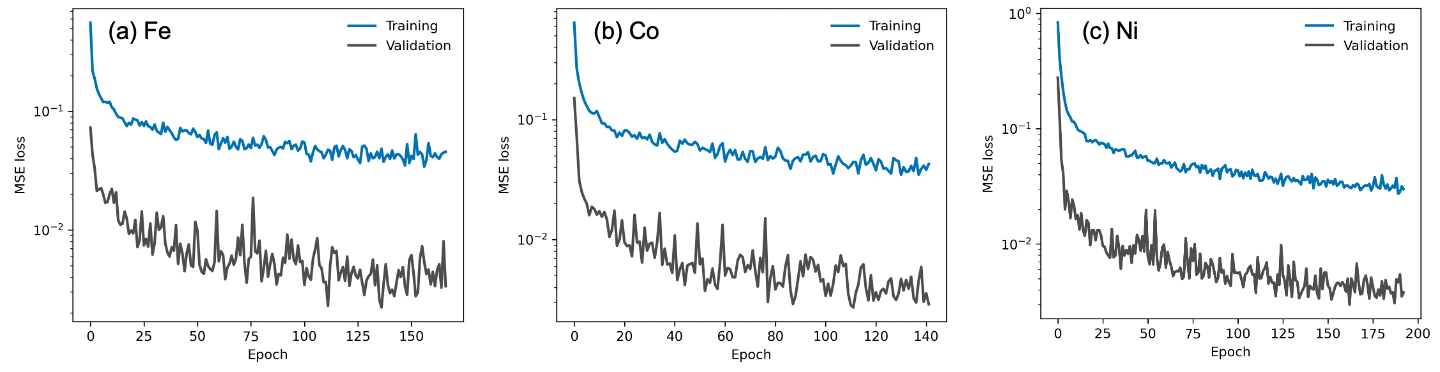}%
\caption{\label{Fig:loss_curves}
Training and validation loss curves of the neural-network model for (a) Fe, (b) Co, and (c) Ni as a function of training epoch.
The mean squared error (MSE) is shown on a logarithmic scale.
}
\end{figure*}

Training and validation loss curves, demonstrating the convergence behavior of the NN, are shown in Fig.~\ref{Fig:loss_curves}.
The validation loss is consistently lower than the training loss, which can be attributed to the use of regularization techniques such as dropout during training, while the validation loss is evaluated without such regularization. 
Both losses decrease smoothly without indications of overfitting.

\begin{figure*}[tb]
\includegraphics[width=\linewidth]{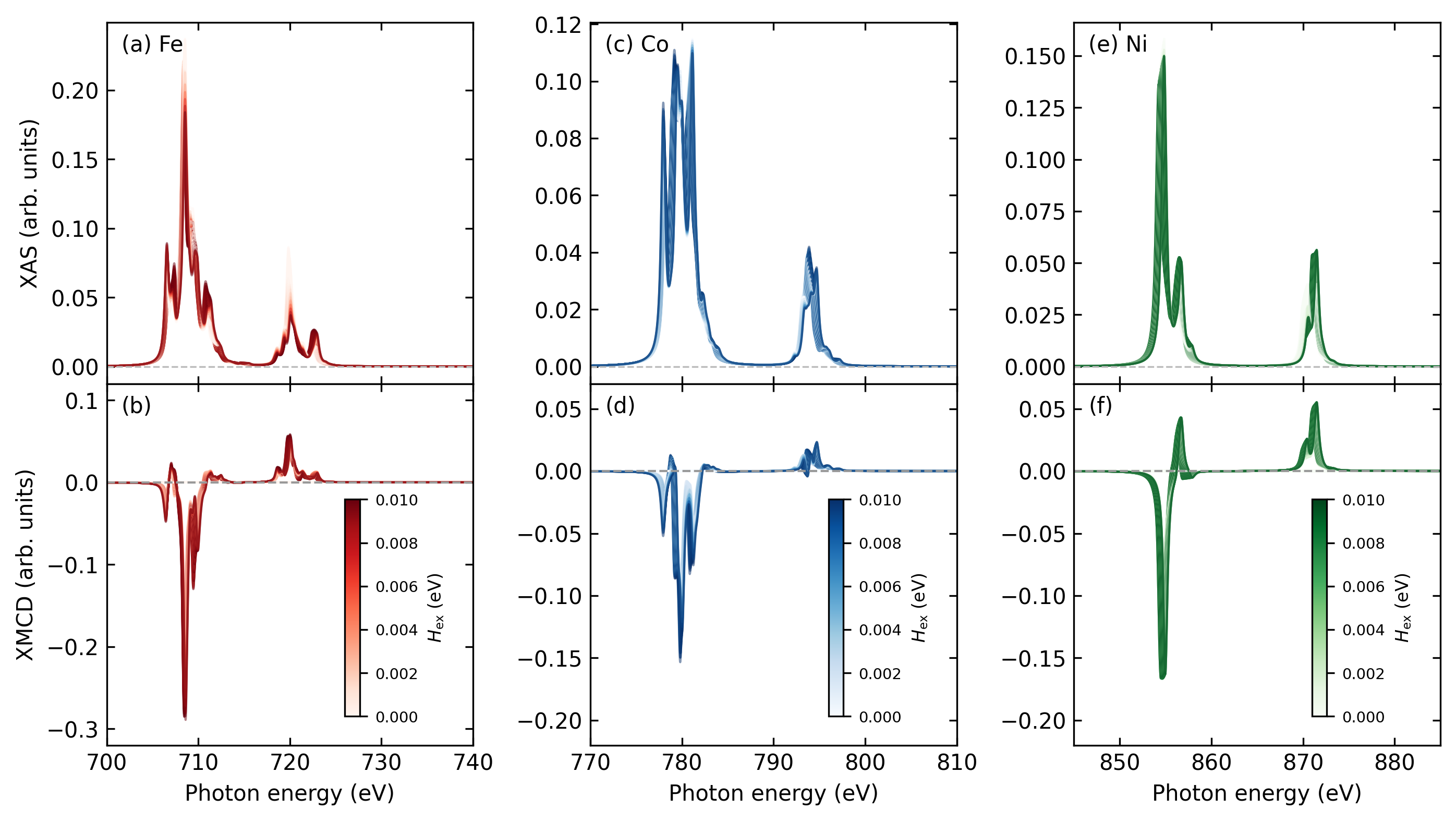}%
\caption{\label{fig:Figure_XAS_XMCD_Hex}
Representative sets of Fe, Co, and Ni $L_{2,3}$ (a) XAS and (b) XMCD spectra calculated using Quanty.
Each panel contains 70 spectra.
The $3d$ electron configurations are assumed to be $3d^6$, $3d^7$ and $3d^8$ for Fe, Co, and Ni, respectively.
The spectra are calculated for different values of the exchange field $H_{\mathrm{ex}}$, while the crystal-field splitting 10Dq is also varied.
All other parameters are fixed at $S_l = 0.70$, $\alpha_{3d} = 1.06$, and $\alpha_{2p} = 0.90$.
}
\end{figure*}

\section{Results}
\label{Results}

\subsection{Parameter dependences of XAS and XMCD spectra}

To elucidate the parameter dependence of Fe, Co, and Ni $L_{2,3}$ XAS and XMCD spectra, Fig.~\ref{fig:Figure_XAS_XMCD_Hex} shows calculated spectra for different values of $H_{\mathrm{ex}}$, while 10Dq is also varied with other parameters fixed. 
Both XAS and XMCD spectra exhibit pronounced features arising from the interplay of crystal-field splitting, spin--orbit coupling, and exchange interaction. In particular, the roles of 10Dq and $H_{\mathrm{ex}}$ can be clearly distinguished.
Variations in 10Dq primarily induce shifts and redistributions of spectral features along the energy axis, reflecting changes in the multiplet structure. In contrast, $H_{\mathrm{ex}}$ mainly affects the magnitude and asymmetry of the XMCD signal, modulating the dichroic intensity without significantly shifting peak positions. 

As a result, the spectral line shapes depend on multiple parameters in a non-trivial manner, with energy-dependent features governed mainly by crystal-field effects and intensity variations controlled by the exchange field.
Such complementary dependencies are not captured by integral-based analyses, highlighting the rich information content of the full spectral line shape and motivating an inverse-analysis approach.


\subsection{Parameter dependences of $\langle S_z \rangle$ and $\langle L_z \rangle$}

\begin{figure*}[tb]
\includegraphics[width=\linewidth]{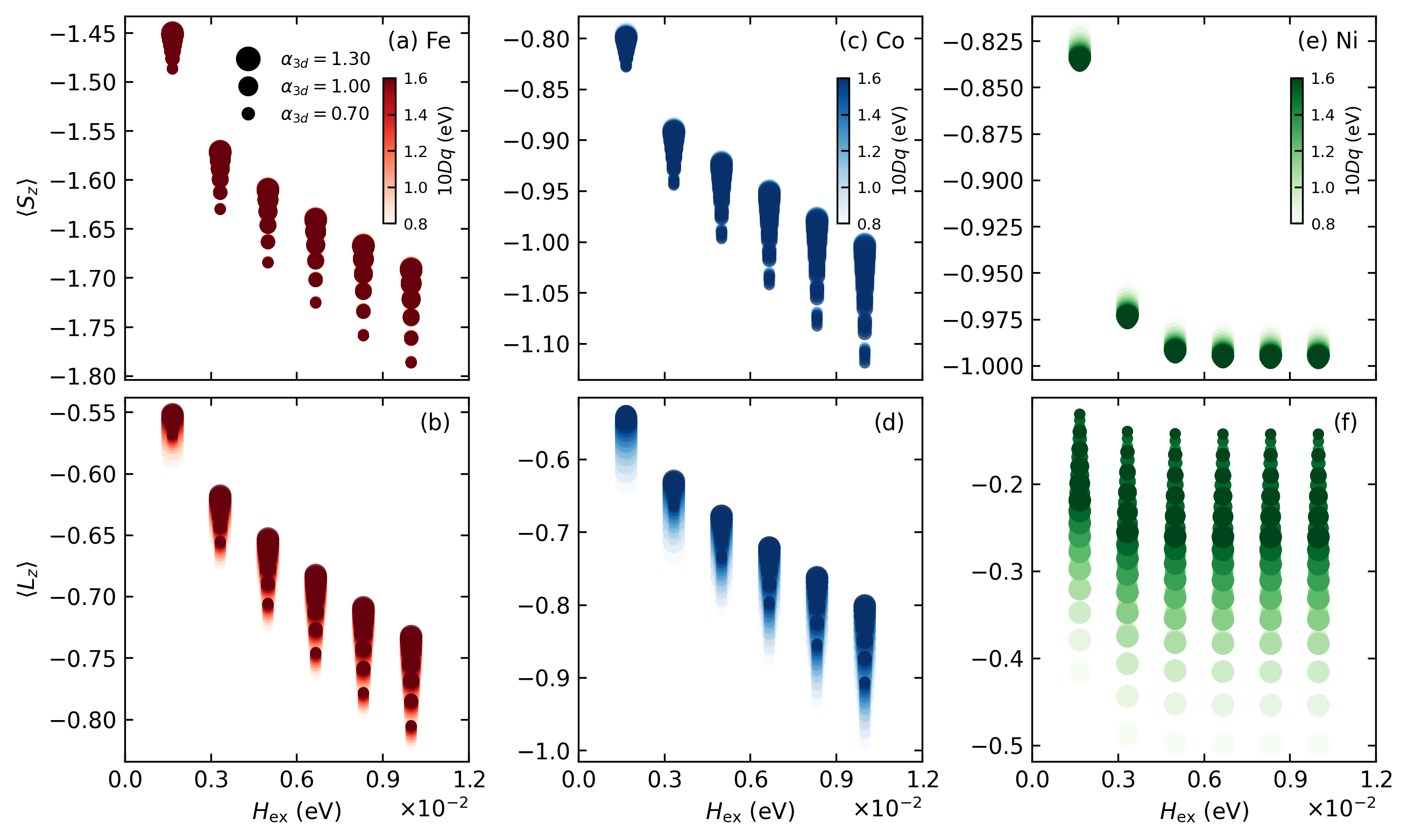}%
\caption{\label{fig:Figure_Sz_Lz_vs_Hex}
Parameter dependences of (a,c,e) $\langle S_z \rangle$ and (b,d,f) $\langle L_z \rangle$ for Fe, Co, and Ni on $H_{\mathrm{ex}}$, 10Dq, and $\alpha_{3d}$.
Color and marker size denote 10Dq and $\alpha_{3d}$, respectively.
Data points with $\langle S_z \rangle = 0$ and $\langle L_z \rangle = 0$ are excluded for consistency with NN training.
}
\end{figure*}

Figure~\ref{fig:Figure_Sz_Lz_vs_Hex} shows the dependence of $\langle S_z \rangle$ and $\langle L_z \rangle$ on $H_{\mathrm{ex}}$, 10Dq, and $\alpha_{3d}$. 
As shown in Figs.~\ref{fig:Figure_Sz_Lz_vs_Hex}(a), \ref{fig:Figure_Sz_Lz_vs_Hex}(c), and \ref{fig:Figure_Sz_Lz_vs_Hex}(e), $|\langle S_z \rangle|$ increases monotonically with increasing $H_{\mathrm{ex}}$, while exhibiting a finite spread at fixed $H_{\mathrm{ex}}$ due to the influence of additional parameters.
Variations in $\alpha_{3d}$ primarily affect the overall magnitude, and 10Dq further modulates the distribution, leading to a multi-valued relationship between $H_{\mathrm{ex}}$ and $\langle S_z \rangle$.
A similar, but more pronounced, multi-parameter dependence is observed for $\langle L_z \rangle$ as shown in Figs.~\ref{fig:Figure_Sz_Lz_vs_Hex}(b), \ref{fig:Figure_Sz_Lz_vs_Hex}(d), and \ref{fig:Figure_Sz_Lz_vs_Hex}(f).
At a given $H_{\mathrm{ex}}$, both 10Dq and $\alpha_{3d}$ significantly influence $\langle L_z \rangle$, resulting in a larger spread than in the case of $\langle S_z \rangle$.
This reflects the stronger sensitivity of the orbital angular momentum to the electronic structure.
In the case of Ni, a weaker dependence of $\langle L_z \rangle$ on $H_{\mathrm{ex}}$ is observed compared with those of Fe and Co.

These results demonstrate that both $\langle S_z \rangle$ and $\langle L_z \rangle$ depend on multiple parameters in a non-trivial manner, consistent with the complex spectral variations shown in Fig.~\ref{fig:Figure_XAS_XMCD_Hex}.
This multi-parameter dependence underscores the necessity of an inverse analysis capable of disentangling these effects.

\subsection{Inverse reconstruction of $\langle S_z \rangle$ and $\langle L_z \rangle$  by neural networks}

\begin{figure*}[tb]
\includegraphics[width=\linewidth]{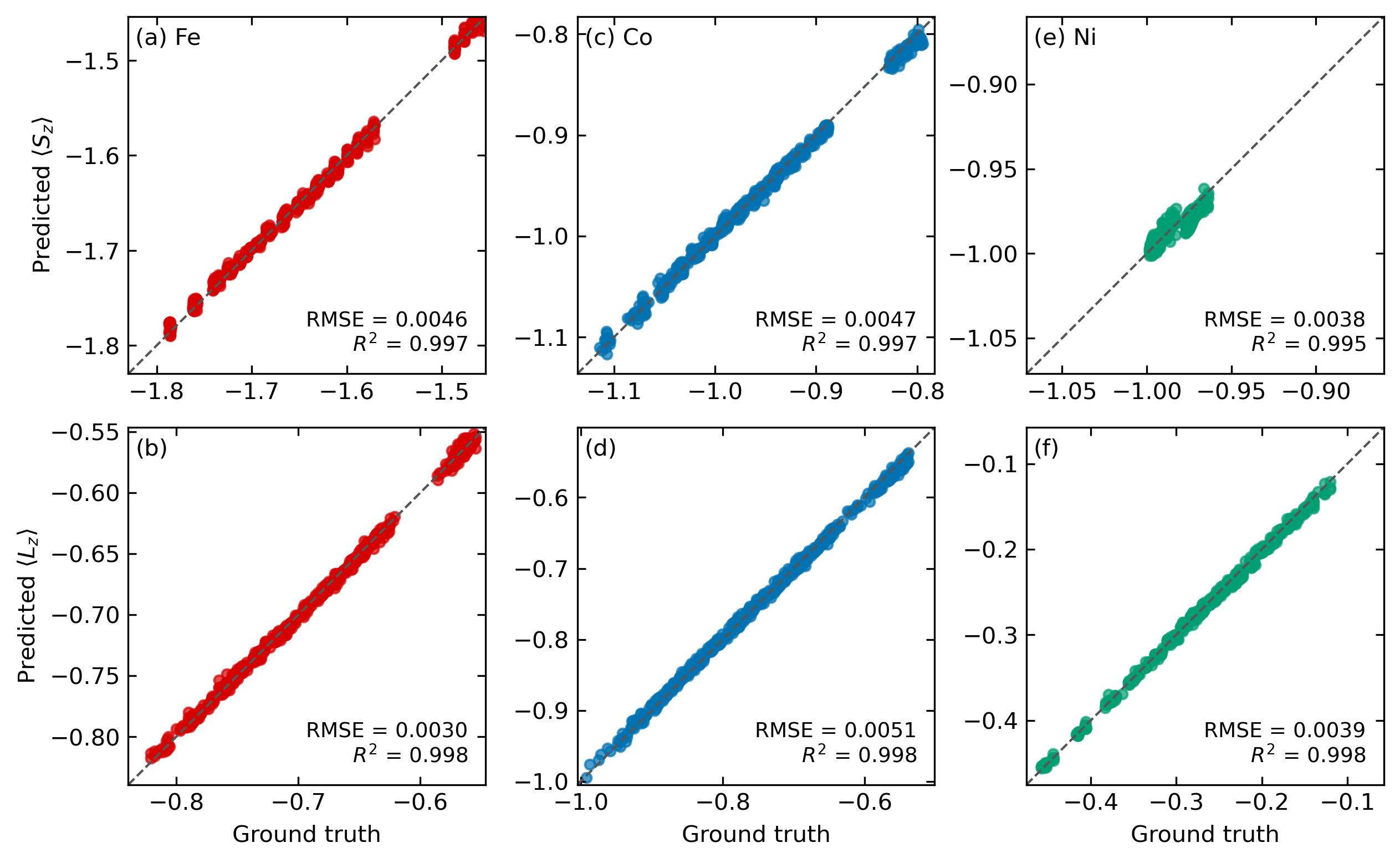}%
\caption{\label{Fig:NNresults_parityplot}
Test-only parity plots for (a,c,e) $\langle S_z \rangle$ and (b,d,f) $\langle L_z \rangle$ for Fe, Co, and Ni comparing the NN predictions with the ground-truth values obtained from the multiplet calculations.
The dashed lines indicate perfect agreement between prediction and ground truth.
The root-mean-square error (RMSE) and the coefficient of determination ($R^2$) are indicated in each panel.
}
\end{figure*}

Here, we show the results of NN-based inverse reconstruction of angular momentum from XAS and XMCD spectra.
Figure~\ref{Fig:NNresults_parityplot} presents test-only parity plots for Fe, Co, and Ni comparing the NN predictions with the ground-truth values obtained from the multiplet calculations.
The predictions closely follow the ideal parity lines over the entire test dataset, indicating that the NN successfully captures the inverse mapping from spectral data to angular-momentum expectation values. 

The reconstruction accuracy is quantified by the root-mean-square error (RMSE) and the coefficient of determination ($R^2$), both of which demonstrate high predictive accuracy without noticeable systematic deviations.
Importantly, all values are reported in physical units after reversing the normalization applied during training.

Because the test dataset is strictly excluded from training and validation, the results reflect genuine generalization rather than memorization.
Together with the results shown in Figs.~\ref{fig:Figure_XAS_XMCD_Hex} and~\ref{fig:Figure_Sz_Lz_vs_Hex}, these findings confirm that the NN enables accurate reconstruction of spin and orbital angular momentum from XAS and XMCD spectra.

\section{Discussion}
\label{Discussion}

The results shown in Fig.~\ref{Fig:NNresults_parityplot} demonstrate that the NN can accurately reconstruct spin and orbital angular momentum from XAS and XMCD spectra when the problem is formulated as an inverse mapping based on full spectral line shapes. 
This capability reflects the fact that the spectra encode information on multiple physical parameters in a distributed and nonlinear manner. 
As shown in Fig.~\ref{fig:Figure_XAS_XMCD_Hex}, crystal-field splitting and exchange field affect the spectra in qualitatively different ways, providing complementary signatures that can be efficiently exploited by the NN.

The reconstruction accuracy $R^2$ for $\langle L_z \rangle$ is marginally higher than that for $\langle S_z \rangle$, although the difference is very small. 
A consistent trend is nevertheless observed across Fe, Co, and Ni suggesting a systematic origin.
This may be attributed to differences in how spin and orbital contributions are encoded in the spectral line shapes. 
In particular, the orbital moment is more directly linked to spin--orbit coupling and anisotropic electronic structure, and is therefore more strongly reflected in detailed spectral features~\cite{Thole_1992}. 
By contrast, the spin contribution is subject to additional ambiguities, such as the magnetic dipole term $T_z$ in the sum-rule framework~\cite{Piamonteze_2009,Sipr_2009}.

It is important to emphasize that the present approach does not replace conventional XMCD sum-rule analyses but complements them. 
While sum rules provide a physically transparent framework based on integrated intensities, the present method exploits the full spectral line shape, enabling access to parameter-dependent features beyond integral-based approaches. 
This advantage is particularly relevant in situations where the assumptions underlying sum rules are not strictly satisfied or where reliable background subtraction is challenging.

Although demonstrated here for Fe, Co, and Ni, the approach is directly applicable to other 3d transition metals and, in principle, to lighter systems such as V, Cr, and Mn, where correction-factor-based analyses are often insufficient~\cite{Goering_2005}.
More broadly, the method provides a flexible framework for analyzing experimental XMCD data, including spatially resolved, operando, and time-resolved measurements.

A potential limitation is that the present dataset is generated within a restricted parameter space, raising the possibility that the task reduces to interpolation. 
However, the mapping from spectral features to angular momentum remains highly nonlinear and involves entangled physical parameters, making the learning problem non-trivial even within this space. 
Moreover, the use of strictly test-only validation confirms that the network captures generalizable physical relationships. 
Nevertheless, the present study is restricted to idealized, noise-free data, and future work should address robustness against noise, background, and other experimental effects.

\section{Conclusion}
\label{Conclusion}

In this work, we formulated the analysis of XAS and XMCD spectra as an inverse problem and demonstrated an NN-based reconstruction of spin and orbital angular momentum from full spectral line shapes. 
Using many-body calculations of Fe, Co, and Ni $L_{2,3}$-edge XMCD spectra as a physically well-defined training dataset, we showed that the network accurately reproduces $\langle S_z \rangle$ and $\langle L_z \rangle$, as confirmed by strictly test-only validation. 
These results demonstrate that exploiting the full XAS and XMCD line shape provides access to information beyond conventional integral-based sum-rule analyses while remaining consistent with established theoretical frameworks. 
Overall, this work establishes a proof of concept for data-driven inverse reconstruction from XAS and XMCD spectra. 
Future work will extend the methodology to more realistic conditions, including energy broadening, noise, and background effects, as well as its application to experimental data.

\begin{acknowledgments}
This work was supported by the Innovative Science and Technology Initiative for Security, Grant Number JPJ004596, ATLA, Japan and JSPS KAKENHI Grant Number JP26K08187.
TU acknowledges the support of the Murata Science and Education Foundation.
\end{acknowledgments}

\section*{Data availability}
The data that support the findings of this study are available from the contact author upon reasonable request.

\clearpage

\section*{APPENDIX: Training and validation loss curves}

\end{document}